\documentclass[aps,preprint,showpacs,floatfix]{revtex4}
\usepackage{graphicx,bm,amsmath,dcolumn}

 \usepackage{graphicx}
 \RequirePackage{color}

 \definecolor{MyDarkGreen}{rgb}{0.02,0.60,0.06}

\graphicspath{{figures}}
\begin{document}
\title{The two-point resistance of fan networks}

\author{N.Sh. Izmailian}
\email{izmail@yerphi.am; ab5223@coventry.ac.uk}
\affiliation{Applied Mathematics Research Centre, Coventry University, Coventry CV1 5FB, UK}
\affiliation{Yerevan Physics Institute, Alikhanian Brothers 2, 375036 Yerevan, Armenia}

\author{R. Kenna}
\email{r.kenna@coventry.ac.uk}
\affiliation{Applied Mathematics Research Centre, Coventry University, Coventry CV1 5FB, UK}

\date{\today}

\begin{abstract}
The problem of the two-point resistance in various networks has recently received considerable attention.
Here we consider the problem on a fan-resistor network, which is a
segment of the cobweb network.
Using a recently developed approach,  we obtain the exact resistance between  two arbitrary nodes on such a network.
As a byproduct, the analysis also delivers the solution of the spanning tree problem on the fan network.
\end{abstract}


\maketitle

\section{Introduction}
\label{Introduction}

In 2004 Wu  derived a general expression for the two-point resistance of a resistor network in terms of the eigenvalues and eigenvectors of the associated Laplacian matrix \cite{wu2004}.
In practice, however, this approach is sometimes difficult to carry through
due to the singular nature of the Laplacian.
In Ref.~\cite{tan2013}, a different method was used to determine the resistances between points at the centre and perimeter of small cobweb networks comprising up to three concentric polygons.
In a recent paper  the current authors, with Wu, revisited the problem of two-point resistance and derived a new and simpler expression  \cite{IKW2014}.
The new expression was then applied to the cobweb resistor network  and the resistance between {\it any} two nodes in a network of {\emph{any}} size was obtained \cite{IKW2014}.
This approach recovered results for small networks obtained in Ref.~\cite{tan2013}.

More recently, Essam, Wu and Tan  considered the fan network, which is the part of the cobweb network, and obtained the resistance between the apex point and a point on the boundary of the network \cite{ETW2014}.
In this paper we apply the approach of Ref.~\cite{IKW2014} to the fan network, and obtain the resistance between {\it any} two nodes in the network.

\section{Resistors on fan networks}
\label{Resistor}
\vskip 0.2 cm
The fan lattice ${\cal L}_{\rm fan}$ is an $M \times N$ lattice of $M$ concentric arcs connected to an apex by $N$ spokes.
It can be considered as a segment of the cobweb network and the example of an $M=3, N=7$ fan network with resistors $s$ and $r$ in the two directions is shown in Fig. 1.
We impose Neumann or free boundary conditions along the two border spokes and along the outermost arc. Sites on  the innermost arc are connected to an external common node (apex).
Therefore there is a total of $M N+1$ nodes.
We use the term Dirichlet-Neumann  to describe the boundary conditions along the innermost apex and outermost arc.

 \begin{figure}[tbp]
  \includegraphics[width=0.8\textwidth]{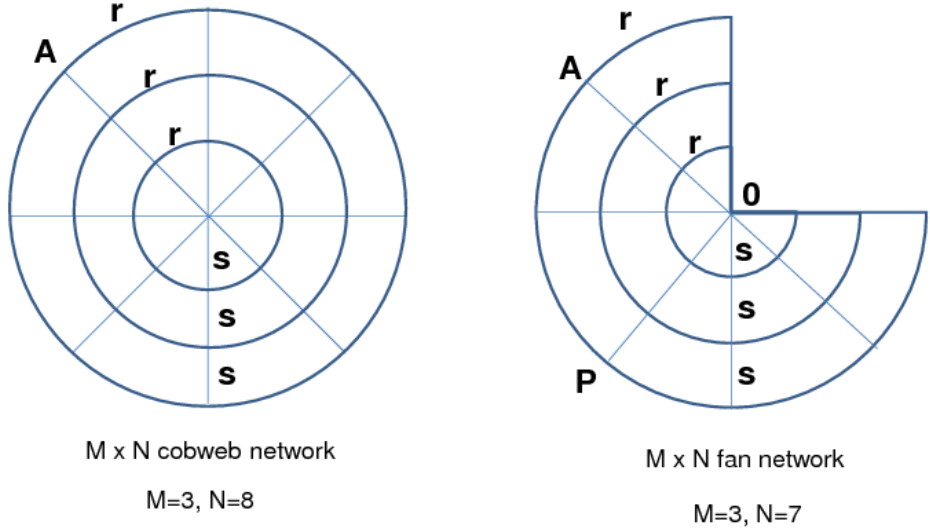}
  \caption{An $M \times N$ fan network with M=3 and N=7 (right). Bonds in the radial and circular directions comprise resistors $s$ and $r$. The apex point is denoted by $O$, $A$ denotes any point on the boundary and $P$
  denotes a point at the middle of the boundary. We  also show $M \times N$ cobweb network with $M=3$ and $N=8$ (left)} \label{fig1}
  \end{figure}

To compute resistances on the fan network, we make use of the formulation given in Ref.~\cite{IKW2014}, and choose the apex node $O$ to be the node $1$ in the fan Laplacian ${\bf L}_{\rm fan}$. This leads us to consider the $(MN) \times (MN)$ cofactor of the $\{1,1\}$-element of ${\bf L}_{\rm fan}$, namely,

\begin{equation}
{\bf \Delta}_{M\times N}=r^{-1}{\bf L}_N^{(\rm{free})}\otimes {\bf I}_M+s^{-1}{\bf I}_N \otimes {\bf L}_M^{(\rm{DN})}\label{Delta}.
\end{equation}
Here, $I_M$ and $I_N$ are  identity matrices and ${\bf L}_N^{(\rm{free})}$ is the Laplacian of a 1d lattice with  free boundary conditions with coordinates between $1$ and $N$, namely
$$
{\bf L}_N^{(\rm{free})}=\left( \begin{array}{ccccccc}
1 & -1 & 0 &\ldots &0&0&0 \\
-1 & 2 & -1&\ldots &0 &0&0\\
\vdots & \vdots & \vdots &\ddots &\vdots & \vdots & \vdots \\
0&0&0&\ldots&-1&2&-1\\
0&0&0&\ldots&0&-1&1
\end{array} \right).
$$
Similarly ${\bf L}_M^{(\rm{DN})}$ is the Laplacian of a 1d lattice with Dirichlet-Neumann boundary conditions and with coordinates between 1 and M, namely
$$
{\bf L}_M^{(\rm{DN})}=\left( \begin{array}{ccccccc}
2 & -1 & 0 &\ldots &0&0&0 \\
-1 & 2 & -1&\ldots &0 &0&0\\
\vdots & \vdots & \vdots &\ddots &\vdots & \vdots & \vdots \\
0&0&0&\ldots&-1&2&-1\\
0&0&0&\ldots&0&-1&1
\end{array} \right).
$$
 The eigenvalues and eigenvectors for ${\bf L}_N^{(\rm{free})}$ and ${\bf L}_M^{(\rm{DN})}$ are known and given by

\begin{eqnarray}
\lambda_n &=& 2 - 2\cos{\theta_n},   \qquad n=0,1,2,...,M-1 , \nonumber\\
f_n(x) &=&\frac{1}{\sqrt{M}}, \qquad n=0 ,\nonumber\\
       &=& \sqrt{\frac{2}{M}}\;\cos{\left((x+1/2)\theta_n\right)}, \qquad n=1,2,...,M-1  ,\nonumber
\end{eqnarray}
for Neumann (free) boundary conditions and given by
\begin{eqnarray}
\lambda_m &=& 2 - 2\cos{(2\varphi_m)},  \qquad m=0,1,2,...,M-1 , \nonumber\\
f_m(y) &=& \frac{2}{\sqrt{2M+1}}\sin{(2 y \varphi_m)},  \qquad m=0,1,2,...,M-1 , \nonumber
\end{eqnarray}
for Dirichlet-Neumann boundary conditions, where
$$
\theta_n=\frac{\pi n}{N} \qquad \mbox{and} \qquad \varphi_{m}=\frac{\pi (m+1/2)}{2M+1}.
$$

This gives the eigenvalues and eigenvectors for the cofactor matrix ${\bf \Delta}_{M \times N}$ of the Laplacian on the fan network as
\begin{eqnarray}
\lambda_{(m,n)}&=&2r^{-1}(1-\cos{\theta_n})+2s^{-1}(1-\cos{2\varphi_m}), \label{lambdamn}\\
\psi_{(m,n);(x,y)}^{cobweb}&=&\frac{2}{\sqrt{N(2M+1)}}\sin{(2y\varphi_m}), \qquad n=0 , \nonumber\\
\psi_{(m,n);(x,y)}^{cobweb}&=&\frac{2\sqrt{2}}{\sqrt{N(2M+1)}}\cos{(\theta_n (x+1/2))}\sin{(2y\varphi_m}),  \qquad n=1,2,...,N-1 \label{psimn} .
\end{eqnarray}
Note, that $\lambda_{(0,0)} \ne 0$.

It follows that the resistance $R^{\rm{fan}}(r_1,r_2)$ between nodes $r_1=(x_1, y_1)$ and $r_2=(x_2,y_2)$ is given by
\begin{eqnarray}
R^{\rm{fan}}(r_1,r_2)&=&\sum_{m=0}^{M-1}\sum_{n=0}^{N-1}
\frac{\left|\psi_{(m,n);(x_1,y_1)}^{\rm{fan}}-\psi_{(m,n);(x_2,y_2)}^{\rm{fan}}\right|^2}
{\lambda_{(m,n)}}\nonumber\\
&=&
\frac{2s}{N(2M+1)}\sum_{m=0}^{M-1}
\left\{{\sum_{n=1}^{N-1}\frac{2(C_1S_1-C_2S_2)^2}
{h(1-\cos{\theta_n})+1-\cos{2\varphi_m}}
+\frac{(S_1-S_2)^2}
{1-\cos{2\varphi_m}}
}\right\},
\label{fangeneral}
\end{eqnarray}
where $h=s/r$ and
\begin{eqnarray}
C_1&=&\cos\left((x_1+1/2)\theta_{n}\right), \qquad C_2=\cos\left((x_2+1/2)\theta_{n}\right), \nonumber\\
S_1&=&\sin\left(2y_1\varphi_m\right), \qquad S_2=\sin\left(2y_2\varphi_m\right). \nonumber
\end{eqnarray}
In particular the resistance between apex node $O=(0,0)$ and a point on the boundary of the network $A=(x,M)$, which we denote as $R^{\rm{fan}}(x)$, is given by
\begin{eqnarray}
R^{\rm{fan}}(x)&=&\frac{s}{N(2M+1)}
\sum_{m=0}^{M-1}
\left\{{
\sum_{n=1}^{N-1}
\frac{2\cos^2{((x+1/2)\theta_n)}\sin^2{(2M\varphi_m)}}
{h \sin^2{\frac{\theta_n}{2}}+\sin^2{\varphi_m}}
+\frac{\sin^2{(2M\varphi_m)}}
{\sin^2{\varphi_m}} }\right\} , \nonumber\\
&=&\frac{s M}{N}+\frac{2s}{N(2M+1)}\sum_{m=0}^{M-1}\sum_{n=1}^{N-1}
\frac{\cos^2{((x+1/2)\theta_n)}\cos^2{\varphi_m}}
{h \sin^2{\frac{\theta_n}{2}}+\sin^2{\varphi_m}},\label{R9}
\end{eqnarray}
with $x \in \{1, \dots N\}$.
Here we have used the fact that
$$
\sin{(2M\varphi_m)}=(-1)^m\cos{\varphi_m}
$$
and the identity
\begin{equation}
\sum_{m=0}^{M-1}\cot^2{\varphi_m} = \sum_{m=0}^{M-1}\cot^2\frac{\pi(m+1/2)}{2M+1}=2M^2+M.
\end{equation}
Using
\begin{equation}
\sum_{n=1}^{N-1}\cos^2{((x+1/2)\theta_n)}=\sum_{n=1}^{N-1}\cos^2{(\pi n (x+1/2)/N)}=\frac{N-1}{2},
\label{identiT}
\end{equation}
for integer $x$, one then obtains
\begin{eqnarray}
R^{\rm{fan}}(x)&=&\frac{s M}{N}-\frac{2s}{N(2M+1)}\sum_{m=0}^{M-1}\sum_{n=1}^{N-1}
\cos^2{((x+1/2)\theta_n)} \left\{{1-\frac{(h \sin^2{\frac{\theta_n}{2}}+1)}
{h \sin^2{\frac{\theta_n}{2}}+\sin^2{\varphi_m}}}\right\}\nonumber\\
&=&
\frac{s M}{N}-\frac{s M(N-1)}{N(2M+1)}
+\frac{2s}{N(2M+1)}\sum_{m=0}^{M-1}\sum_{n=1}^{N-1}
\frac{\cos^2{((x+1/2)\theta_n)}(h \sin^2{\frac{\theta_n}{2}}+1)}
{h \sin^2{\frac{\theta_n}{2}}+\sin^2{\varphi_m}}.~~~~~\label{R10}
\end{eqnarray}
The summation over $m$ in Eq. (\ref{R10}) can be extended up to $2M$ as follows
\begin{eqnarray}
R^{\rm{fan}}(x)
&=&\frac{s M}{N}-\frac{s M(N-1)}{N(2M+1)}-\frac{s}{N(2M+1)}\sum_{n=1}^{N-1}
\cos^2{((x+1/2)\theta_n)}\nonumber\\
&+&\frac{s}{N(2M+1)}\sum_{n=1}^{N-1}\sum_{m=0}^{2M}
\frac{\cos^2{((x+1/2)\theta_n)}(h \sin^2{\frac{\theta_n}{2}}+1)}
{h \sin^2{\frac{\theta_n}{2}}+\sin^2{\varphi_m}}\label{R11}\\
&=&\frac{s M}{N}-\frac{s (N-1)}{2N}+\frac{s}{N(2M+1)}\sum_{n=1}^{N-1}\sum_{m=0}^{2M}
\frac{\cos^2{((x+1/2)\theta_n)}(h \sin^2{\frac{\theta_n}{2}}+1)}
{h \sin^2{\frac{\theta_n}{2}}+\sin^2{\varphi_m}},\label{R12}
\end{eqnarray}
having again used Eq. (\ref{identiT}).

Now using the identity (see for example  Eq. (27) of Ref.~\cite{izmailian2010})
\begin{equation}
\sum_{m=0}^{2M} \Big[{\sin^2 \frac{\pi (m+1/2)}{2M+1} +h\sin^2
\frac{\pi n}{2N}}\Big]^{-1}=2 (2M+1) \frac{{\rm
tanh}\left[(2M+1)\,\omega(\frac{\pi n}{2N})\right]}{{\rm \sinh
2\omega(\frac{\pi n}{2N})}} , \nonumber
\end{equation}
where $\omega(x)$ is given by
\begin{equation}
\omega(x) = {\rm arcsinh}\sqrt{h} \sin x \label{omega} ,
\end{equation}
we  finally arrive at
\begin{eqnarray}
R^{\rm{fan}}(x)&=&\frac{s (2M+1-N)}{2N}+\frac{2s}{N}\sum_{n=1}^{N-1}
\frac{\cos^2{((x+1/2)\theta_n)}(h \sin^2{\frac{\theta_n}{2}}+1)}
{\rm \sinh
2\omega(\frac{\theta_n}{2})}{\rm
tanh}\left[(2M+1)\,\omega\left(\frac{\theta_n}{2}\right)\right] \nonumber\\
&=&\frac{s (2M+1-N)}{2N}+\frac{s}{N}\sum_{n=1}^{N-1}
\cos^2{((x+1/2)\theta_n)}\frac{{\rm
tanh}\left[(2M+1)\,\omega\left(\frac{\theta_n}{2}\right)\right]}{{\rm
tanh}\left[\omega\left(\frac{\theta_n}{2}\right)\right]}.\label{Rfin}
\end{eqnarray}

From Eq. (\ref{Rfin}) one can see that $R^{\rm{fan}}(x)$ has the following symmetry
\begin{equation}
R^{\rm{fan}}(x)=R^{\rm{fan}}(N-1-x).
\end{equation}

For $N$ odd let us calculate the value of $R^{\rm{fan}}(x)$ at $x=(N-1)/2$, which is middle point of the boundary of the fan network considered in Ref.~\cite{ETW2014}.
The resistance between the apex node $O=(0,0)$ and this middle point $P=(\frac{N-1}{2}, M)$ can be obtained from Eq. (\ref{Rfin}) in the  form
\begin{eqnarray}
R^{\rm{fan}}\left(\frac{N-1}{2}\right)&=&\frac{s (2M+1-N)}{2N}+\frac{s}{N}\sum_{n=1}^{N-1}
\cos^2{(N\theta_n/2)}\frac{{\rm
tanh}\left[(2M+1)\,\omega\left(\frac{\theta_n}{2}\right)\right]}{{\rm
tanh}\left[\omega\left(\frac{\theta_n}{2}\right)\right]} \nonumber\\
&=&\frac{s (2M+1-N)}{2N}+\frac{s}{N}\sum_{n=1}^{\frac{N-1}{2}}
\frac{{\rm
tanh}\left[(2M+1)\,\omega\left(\theta_n\right)\right]}{{\rm
tanh}\left[\omega\left(\theta_n\right)\right]}.\label{R92}
\end{eqnarray}
Here we use that $\cos^2{(N\theta_n/2)}=0$ for odd n and equals 1 for even n. Using the symmetry property of the $\omega(x)$
\begin{equation}
\omega{(\pi-x)}=\omega{(x)}
\nonumber
\end{equation}
we can extend the summation over $n$ in Eq. (\ref{R92}) from $(N-1)/2$ up to $N-1$ and obtain the following expression
\begin{eqnarray}
R^{\rm{fan}}\left(\frac{N-1}{2}\right)&=&\frac{s (2M+1-N)}{2N}+\frac{s}{2N}\sum_{n=1}^{N-1}
\frac{{\rm
tanh}\left[(2M+1)\,\omega\left(\theta_n\right)\right]}{{\rm
tanh}\left[\omega\left(\theta_n\right)\right]} \label{Rfancobweb}.
\end{eqnarray}

\section{Connection with  the results of Essam, Tan and Wu.}
\label{Resistor1}
In this section we will show that our results for the resistance between two particular points, namely between apex $O=(0,0)$ and a point on the boundary $A=(x,M)$, coincides with results of Essam, Tan and Wu \cite{ETW2014}.
The connection is made by summing over $m$ instead of over $n$ as was done in Sec.II.

Let us start with Eq. (\ref{R9}), which can be transformed as
\begin{eqnarray}
R^{\rm{fan}}(x)
&=&\frac{s M}{N}+\frac{r}{N(2M+1)}\sum_{m=0}^{M-1}\sum_{n=1}^{N-1}
\frac{\left[1+\cos{((2x+1)\theta_n)}\right]\cos^2{\varphi_m}}
{\sin^2{\frac{\theta_n}{2}}+h^{-1}\sin^2{\varphi_m}} \label{R9bb}\\
&=&-\frac{s M}{N}+\frac{2r}{N(2M+1)}\sum_{m=0}^{M-1}\sum_{n=0}^{N-1}
\frac{\left[1+\cos{((2x+1)\theta_n)}\right]\cos^2{\varphi_m}}
{\cosh{(2\Lambda_m)}-\cos{\theta_n}},\label{R101}
\end{eqnarray}
where $\Lambda_m$ is defined as
\begin{equation}
\sinh\Lambda_m=\sqrt{h^{-1}}\sin{\varphi_m}\label{Lambda}.
\end{equation}
Note, that we have changed the second summation in Eq. (\ref{R101})
to start at $n=0$ instead of at $n=1$.

We can then carry out the summation over $n$ in (\ref{R101}) by using the summation identities (see Eq. (61) of Ref. \cite{wu2004} with $\lambda=2 \Lambda$)
\begin{equation}
\frac{1}{N} \sum_{n=0}^{N-1} \frac {\cos (\ell \theta_n)}
 {\cosh (2 \varLambda) - \cos \theta_n   } = \, \frac{\cosh[2(N - \ell)\varLambda)]}
{\sinh (2\varLambda )\sinh(2N \varLambda)}+\frac{1}{N}\left[\frac{1}{\sinh^2 (2\varLambda )}+\frac{1-(-1)^l}{4\cosh^2 \varLambda}\right],
\label{sumidentity}
\end{equation}
with $ \ell = 0, \> 2x+1$,
to obtain
\begin{eqnarray}
R^{\rm{fan}}(x)
&=&-\frac{s M}{N}+\frac{2r}{(2M+1)}\sum_{m=0}^{M-1}\frac{\cosh (2 N \varLambda_m)+\cosh (2(N-2x-1)\varLambda_m)}{\sinh(2 \varLambda_m)\sinh(2N \varLambda_m)}\cos^2{\varphi_m}\nonumber\\
&+&\frac{r}{N(2M+1)}\sum_{m=0}^{M-1}\frac{\cos^2{\varphi_m}}{\sinh^2\Lambda_m} 
\label{R102}\\
&=&\frac{2r}{(2M+1)}\sum_{m=0}^{M-1}\frac{\cosh (2 N \varLambda_m)+\cosh (2(N-2x-1)\varLambda_m)}{\sinh(2 \varLambda_m)\sinh(2N \varLambda_m)}\cos^2{\varphi_m}.\label{R103}
\end{eqnarray}
Here we have used the identity
\begin{equation}
\sum_{m=0}^{M-1}\frac{\cos^2{\varphi_m}}{\sinh^2\Lambda_m}=h\sum_{m=0}^{M-1}\cot^2{\varphi_m}=M(2M+1)h.
\label{hoover}
\end{equation}
Eq.~(\ref{R103}) is  Eq. (3.11) of Ref. \cite{ETW2014} with $i=m+1$, $t=x$ and $\lambda_m=e^{2\Lambda_m}$.

The resistance between the apex node $O=(0,0)$ and the middle point $P=(\frac{N-1}{2}, M)$ can be obtained for odd $N$ from Eq. (\ref{R103}) in the   form
\begin{eqnarray}
R^{\rm{fan}}\left(\frac{N-1}{2}\right)
&=&\frac{2r}{(2M+1)}\sum_{m=0}^{M-1}\frac{\coth (2 N \varLambda_m)}{\sinh(2 \varLambda_m)}\cos^2{\varphi_m},\label{Rmiddle}
\end{eqnarray}
which is exactly the resistance between central node and the boundary node on the cobweb network having the same number $N$ of radial lines
as identified in  Ref.~\cite{ETW2014}.
The results  (\ref{R103}) and (\ref{Rmiddle}), which first appeared in Ref.~\cite{ETW2014}, are particular cases of the general result (\ref{fangeneral}).

\section{Spanning tree on fan networks}
\label{Spanning}
As a byproduct of our analysis, we solve the problem of enumerating weighted spanning trees on
an $M\times N$  fan network ${\cal L}_{{\rm{fan}}\ M\times N}$.

The problem of enumerating spanning trees on a graph was first considered by Kirchhoff in his analysis of electrical networks \cite{Kirchhoff}. The enumeration of spanning trees concerns the evaluation of the tree generating function
\begin{equation}
Z_{{\rm{fan}}\ (M\times N)}^{\rm Sp}(x,y)=\sum_T x^{n_x}y^{n_y}
\label{span}
\end{equation}
where we assign weights $x$ and $y$, respectively, to edges in the spokes and concentric arcs, and the summation is taken over all spanning tree configurations $T$ on
${\cal L}_{ {\rm{fan}}\ (M\times N)}$ with $n_x$ and $n_y$ edges  in the respective directions. Setting $x=y=1$ we obtain
\begin{equation}
Z_{{\rm{fan}}\ (M\times N)}^{\rm Sp}(1,1)=\mbox{the number of spanning trees on the fan network}.
\label{Nspan}
\end{equation}

It is well known \cite{Brooks,Harary,tzengwu1} that the spanning-tree generating function is given by the determinant of the cofactor of {\it any} element of the Laplacian matrix of the network.
We can therefore  evaluate ${\bf \Delta}_{MN}$ given in (\ref{Delta}) with $r^{-1}=x, s^{-1}=y$. This gives
\begin{eqnarray}
Z_{{\rm{fan}}\ (M\times N)}^{\rm Sp}(x,y) &=& \det |{\bf \Delta}_{MN}| \nonumber \\
                         &=& \prod_{m=0}^{M-1}\prod_{n=0}^{N-1}\Lambda_{m,n}(x,y),
\label{span2}
\end{eqnarray}
where $\Lambda_{m,n}(x,y)$ is given by Eq. (\ref{lambdamn}) with $r^{-1}= x$ and $s^{-1} =y$. Thus, we obtain the closed form expression for the spanning tree generating function
\begin{eqnarray}
 Z_{{\rm{fan}}\ (M\times N)}^{\rm Sp}(x,y) &=& \prod_{m=0}^{M-1}\prod_{n=0}^{N-1}\left[2x\left(1-\cos{\frac{\pi n}{N}}\right)+2y\left(1-\cos{\frac{\pi(2m+1}{2M+1}}\right)\right]
\nonumber \\
&=&\prod_{m=0}^{M-1}\prod_{n=0}^{N-1}4\left[x \sin^2{\frac{\pi
n}{2N}}+y\sin^2{\frac{\pi(m+\frac 1 2 )}{2M+1}}\right]\label{spangen2}.
\end{eqnarray}

In comparison, the spanning tree generating function for an $M \times N$ plane lattice with free boundary conditions in $N$ and $M$ directions computed by Tzeng and Wu \cite{tzengwu1} is
 \begin{equation}
Z_{{\rm{plane}}\ (M\times N)}^{\rm Sp}(x,y) = \frac{1}{M N} \prod_{m=0}^{M-1}
\prod_{n=0 \above0pt  (m,n) \neq (0,0)}^{N-1} \left[2x\left(1-\cos{\frac{ n
\pi}{N}}\right)+2y\left(1-\cos{\frac{m
\pi}{M}}\right)\right].
\label{spancyl1}
 \end{equation}
The expression (\ref{spancyl1}) can be transformed to
\begin{equation}
Z_{{\rm{plane}}\ (M\times N)}^{\rm Sp}(x,y)=x^{N-1}y^{M-1}
\prod_{m=1}^{M-1}\prod_{n=1}^{N-1} 4\left[x \sin^2{\frac{\pi
n}{2N}}+y\sin^2{\frac{\pi m}{2M}}\right], \label{spancylfin}
\end{equation}
by using the identities
\begin{equation}
\prod_{k=1}^{P-1} 4 z \sin^2 \frac {\pi k} { 2 P} = P\,  z^{P-1}  .\nonumber
\end{equation}
The expression (\ref{spancylfin}) can now be compared to (\ref{spangen2}) for the $M\times N$ fan lattice.
Particularly, for $M=3, N=7$,
we obtain for the $3\times 7$ fan lattice the number
 \begin{equation}
 Z_{{\rm{fan}}\ (3\times 7)}^{\rm Sp}(1,1)= 536\ 948\ 224,  \nonumber
\end{equation}
and for the  $3\times 7$ plane lattice the number
\begin{equation}
Z_{{\rm{plane}}\ (3\times 7)}^{\rm Sp}(1,1)= 4\ 768\ 673. \nonumber
\end{equation}
 The addition of one apex node to a $3\times 7$ plane lattice increases the number of spanning trees
by more than 100 times!

\section{Summary}
\label{Summary}
The method of Izmailian, Kenna and Wu \cite{IKW2014} has been used to derive the resistance between two {\emph{arbitrary}}  nodes of fan networks.
This general result is given in Equation~(\ref{fangeneral}).
From this general result, the resistance between the apex point $O$ and any point $A$ on the perimeter, at distance $M$ from $O$, follows and is given by  Eq.~(\ref{Rfin}) or Eq.~(\ref{R103}).
The symmetric case where $A$ is equidistant from the corner points is given by Eq.~(\ref{Rfancobweb}) or Eq.~(\ref{Rmiddle}).
These recover particular results of Ref.~\cite{ETW2014}.
The solution of the spanning tree problem on fan networks follows as Eq.~(\ref{spangen2}), a byproduct of the above results.


\begin{thebibliography}{99}

\bibitem{wu2004} F.Y. Wu, J. Phys. A: Math. Gen. {\bf 37}, 6653 (2004).

\bibitem{tan2013} Z.-Z. Tan, L. Zhou and J.-H. Yang, J. Phys. A: Math. Theor. {\bf 46}, 195202 (2013).


\bibitem{IKW2014} N.Sh. Izmailian, R. Kenna and F.Y. Wu, J. Phys. A: Math. Theor. {\bf 47}, 035003 (2014).

\bibitem{ETW2014} J.W. Essam, Zhi-Zhong Tan and F.Y. Wu, {\sl Proof and extension of the resistance formula for an
m x n fan network conjectured by Tan, Zhou and Yang}, arXiv:1312.6727.







\bibitem{izmailian2010} N.Sh. Izmailian and M.-C. Huang, Phys. Rev. E {\bf 82}, 011125 (2010).


\bibitem{Kirchhoff} G. Kirchhoff, Ann. Phys. und Chemie. {\bf 72}, 497 (1847).
\bibitem{tzengwu1} W.J. Tseng and F.Y. Wu, Appl. Math. Lett. {\bf 13}, 19 (2000).
\bibitem{Brooks} R.L. Brooks, C.A.B. Smith, A.H. Stone and W.T. Tutte, Duke Math. J. {\bf 7}, 312 (1940).

\bibitem{Harary} F. Harary, {it Graph Theory}, Addison-Wesley, Reading, MA, (1969).

\end{thebibliography}
\end{document}